\documentclass[twocolumn,aps,prl,superscriptaddress,floatfix]{revtex4-1}
\usepackage{graphicx}
\usepackage{amsmath}
\usepackage{tabularx}
\usepackage[caption=false]{subfig}
\usepackage{epstopdf}

\begin{document}
\title{Itinerant-localized model of strongly correlated electrons: \\Fermi-surface reconstruction}

\author{Ilya Ivantsov}
\affiliation{Bogoliubov Laboratory of Theoretical Physics, Joint
Institute for Nuclear Research, Dubna, Russia}
\affiliation{L.V.Kyrensky Institute of Physics, Siberian Branch of Russian Academy of Sciences, Krasnoyarsk, Russia}
\author{Alvaro Ferraz}
\affiliation{International Institute of Physics - UFRN,
Department of Experimental and Theoretical Physics - UFRN, Natal, Brazil}
\author{ Evgenii Kochetov}
\affiliation{Bogoliubov Laboratory of Theoretical Physics, Joint
Institute for Nuclear Research, Dubna, Russia}

\begin{abstract}

A number of recent experiments have highlighted a remarkable transformation of a large cuprate Fermi surface into small pockets in the underdoped region signalling a breakdown of a conventional Fermi liquid theory in the PG phase.
A few phenomenological models have been recently put forward to account for this transformation. However, none of those models
have been derived microscopically nor are totally compatible with experimental data. In the present work we show that the observed Fermi-surface reconstruction can be accounted for directly within a standard microscopic $t-J$ model of correlated electrons, provided strong electron correlations are properly taken into account.

\end{abstract}
%\pacs x 74.20.Mn, 74.20.-z
\maketitle

\section{I. Introduction} The experimental observation of quantum oscillations (QO) in the lightly hole-doped cuprates in high magnetic field \cite{DL} is an important breakthrough
since it indicates that coherent quasiparticles might exist even in the PG regime. The theory does not exhibit a large Fermi surface (FS) enclosing the total number
of charged carriers. Instead, the FS consists of small pockets with a total area
proportional to the dopant density $\delta$, rather than the $1+\delta$ which is expected for conventional Fermi liquids (FLs).

This remarkable crossover between those two regimes occurs at a hole density around $\delta_c\approx 0.19.$\cite{he} Both Hall and Seebeck coefficients are negative at low temperature in high magnetic fields below $\delta_c.$
This indicates that, despite the hole doping, the FS reconstructs into small {\it electronlike} pockets
dominating the transport at low temperature. All translational symmetry-breaking models proposed so far predict however {\it hole} pockets (or open "arcs") at the nodal point $(\frac{\pi}{2}, \frac{\pi}{2})$ in the Brillouin zone (BZ)
and the possibility of an electron pocket at the antinodal point $(\pi,0)$. The latter option seems unphysical since it is
rather unlikely that antinodal pockets and the PG state might exist simultaneously at the same location in the BZ.
The more exotic route involves a new metallic state -- the so-called fractionalized Fermi liquid (FL$^*$)
-- which exhibits small hole pockets
similar to what is observed in an antiferromagnetic (AF) metal, keeping at the same time the translational symmetry intact.\cite{frac} However, the necessary high-temperature remnant ingredients for such an approach -- the associated $Z_2$ topological gauge excitations "visons" -- have never been detected in the cuprates so far.

The most promising approach to account for this FS reconstruction into small electronlike pockets is presumably the one
which assumes that it results from the charge density wave (CDW) modulation observed in the PG phase for hole doping levels $\delta_{c1}\le \delta\le \delta_{c2}.$
The onset temperature of the CDW correlations forms a "dome" ranging from $\delta_{c1}\sim 0.08$ to $\delta_{c2}\sim 0.16$ in the $\delta-T$ phase diagram, with a peak of $T_{CDW}\sim 160K$ for $\delta \sim 0.12.$\cite{RXS} This is also where the QO amplitude is maximal. As a matter of fact Harrison and Sebastian \cite{HS} proposed a phenomenological model in which nodal electron pockets can be understood as a consequence of a bidirectional CDW order. This may provide a way to explain the small QO observed in YBCO, as well as the negative Hall and Seebeck coefficients observed at low temperature in high magnetic fields. Note that those pockets are obtained in the presence of the long-range CDW order. At zero magnetic field the order is short-range and the effect is less visible
by a zero field alternative to QO, the angle resolved photoemission spectroscopy (ARPES). The sides of the pocket manifest themselves as disconnected arc-like features near the nodal region. The onset of PG is defined by the opening of an anti-nodal gap and reconstruction of the large FS to a small pocket or a Fermi "arc" which may actually be one side of a Fermi pocket. Since it occurs approximately at $\delta =\delta_{c}> \delta_{c2}$, the CDW correlations should not be considered as the root cause of the PG phenomenon. It rather appears as just an instability inside the still mysterious PG phase.\cite{RXS}

In spite of its appeal the approach based on the CDW modulations to explain the electron-like nature of the small FS pockets exposed in \cite{HS, sachdev}
remains to this date on a phenomenological level
that cannot be justified within a microscopic framework of the $t-J$ model at the relevant values for the incoming parameters.
Besides, it is
still unclear whether the small FS is indeed merely revealed by QO or possibly created by the necessary high magnetic field.
A natural question then arises of whether or not the FS reconstruction observed in the underdoped cuprates can be accounted for directly within the standard $t-J$ model at zero field. The main aim of the present work is to demonstrate that this is indeed the case, provided the strong electron correlations encoded into the local
no double occupancy (NDO) constraint are properly taken into account.

\section{II. Itinerant-localized model} In the underdoped cuprates, one striking feature is the simultaneous existence of both
localized and itinerant nature of the lattice electrons.
The interplay between spin and charge degrees of freedom plays a dominant role in determining the physics of strongly correlated electrons.
It is therefore convenient to rewrite the $t-J$ model Hamiltonian in a form well suited to take these degrees of freedom explicitly into account.
One might hope that representing the $t-J$ model in a form that takes
both aspects into consideration, on an equal footing, would help to address the problem in a more efficient way both analytically and numerically.

For this purpose Ribeiro and Wen proposed a slave-particle spin-dopon representation of the projected electron operators
in an enlarged Hilbert space \cite{wen},
\begin{eqnarray}
\tilde c_{i}^{\dagger}
=c_{i}^{\dagger}(1-n_{i-\sigma})=\frac{1}{\sqrt{2}}(\frac{1}{2}-2\vec S_i\cdot\vec\sigma)\tilde d_i.
\label{1.1}\end{eqnarray}
In this framework, the localized feature of the electron is represented by the lattice spin $\vec S\in su(2)$
whereas its itinerant nature is described by the doped hole (dopon) represented by the projected hole operator,
$\tilde{d}_{i\alpha}=d_{i\sigma}(1-n^d_{i-\alpha})$.
Here $\tilde c^{\dagger}=(\tilde c^{\dagger}_{\uparrow},\tilde c^{\dagger}_{\downarrow})^{t}$ and
$\tilde d=(\tilde d_{\uparrow}, \tilde d_{\downarrow})^{t}$.

The NDO constraint to encode strong correlations reduces to a Kondo-type interaction,\cite{pfk}
\begin{equation}
\sum_{\alpha}(\tilde c_{i\alpha}^{\dagger}\tilde c_{i\alpha})+\tilde c_{i\alpha}\tilde c_{i\alpha}^{\dagger}-1=
\vec{S_i} \cdot
\vec{s_i}+\frac{3}{4}(\tilde{d}_{i\uparrow}^{\dagger}\tilde{d}_{i\uparrow}+
\tilde{d}_{i\downarrow}^{\dagger}\tilde{d}_{i\downarrow})=0,
\label{a} \end{equation}
with $\vec
s_i=\sum_{\alpha,\beta}\tilde{d}_{i\alpha}^{\dagger}\vec\sigma_{\alpha\beta}\tilde{d}_{i\beta}
$ being the dopon spin operator.
As a result, the canonical $t-J$ model can be reduced to
a lattice Kondo-Heisenberg model at a dominatingly large Kondo coupling to describe the underdoped cuprates,\cite{ifk}
\begin{eqnarray}
H_{t-J}&=& \sum_{ij\sigma} 2t_{ij}
{d}_{i\sigma}^{\dagger} {d}_{j\sigma}
+ J\sum_{ij}(\vec S_i\cdot \vec S_j)\nonumber\\
&+&\lambda\sum_i(\vec{S_i} \cdot
\vec{s_i}+\frac{3}{4}n^d_i),\quad \lambda/t\gg 1.
\label{1.7}\end{eqnarray}

In spite of the global character of the parameter $\lambda$, it enforces the NDO constraint locally due to the fact that
the on-site physical Hilbert subspace corresponds to zero eigenvalues of the constraint, whereas the nonphysical subspace
is spanned by the eigenvectors with strictly positive eigenvalues.
The unphysical doubly occupied electron states are separated from the physical
sector by an energy gap $\sim\lambda$.
In the $\lambda\to +\infty$ limit, i.e. in the limit in which $\lambda$ is much larger than any other existing energy
scale in the problem, those states
are automatically excluded from the Hilbert space so that our model represents the
canonical $t-J$ model.
At a finite $\lambda$ the NDO constraint is relaxed, and as soon as this happens the unwanted unphysical states are allowed to contribute to the theory as well.
In the $\lambda\to 0$ limit, no correlations are present and the problem reduces to its weak
coupling limit that exhibits a conventional Fermi liquid behaviour.

\section{III. Method} We employ the Cluster Perturbation Theory (CPT) method described in
\cite{maier,senechal-prb,senechal-prl} to calculate the spectral function $A({\bf
k},\omega)=-\frac{1}{\pi}Im G({\bf k},\omega)$,
where $G({\bf k},\omega)$ is the single-particle Green function at momentum ${\bf
k}$ and energy $\omega$.
We focus our attention on the spectral function properties
at $\omega=0$ which are then directly related to a relevant FS.

The CPT scheme is based upon an exact diagonalization of small clusters and on
the accountability of an intercluster interaction by means of an appropriate perturbation theory.
We define the Hubbard operators $X^{pq}_I=|p\rangle_I\langle q|_I$ where the $|p\rangle_I$ and $\langle q|_I$ are the eigenstates of the $I^{th}$ cluster (from now on the capital letters stand for the cluster indexes).
In this basis the Hamiltonian (\ref{1.7}) takes on the form:
\begin{equation}
H_{t-J}=\sum_{Ip}\varepsilon_pX^{pp}_I+\sum_{IJ}\sum_{\alpha\beta}T_{IJ}^{\alpha\beta}X_I^{\alpha}X_J^{\beta},
\end{equation}
where $\varepsilon_p$ is the $p$-th eigenvalue and the $\alpha$ and $\beta$ denote composite indexes, e.g., $\alpha=(p,q)$.
The first diagonal term accounts for the interaction within the $I^{th}$ cluster.
The second term describes the intercluster interaction.
The $T_{IJ}^{\alpha\beta}$ are hopping integrals between clusters.
The Fourier transform of the Green function, $D_{\alpha\beta}(\tilde{\bf k},\omega)=\ll X_{\tilde{\bf k}}^{\alpha}| X_{\tilde{\bf k}}^{\beta}\gg_{\omega}$, is taken in the form \cite{senechal-prb}
\begin{equation}
D(\tilde{\bf k},\omega)^{-1}=D^0(\omega)^{-1}-T(\tilde{\bf k}),
\end{equation}
where $$D^0_{\alpha\beta}(\omega)=\delta_{\alpha\beta}\frac{\langle X^{pp}\rangle_0+\langle X^{qq}\rangle_0}{\omega-\varepsilon_q+\varepsilon_p+\mu}.$$ Here $\mu$ is the chemical potential and the $\tilde{\bf k}$ is
a corresponding wave vector defined in the reduced Brillouin zone \cite{maier}.
The full-lattice electron Green function to calculate spectra reads
\begin{equation}
G_{\sigma}({\bf k},\omega)=\frac{1}{N_c}\sum_{\alpha\beta}\sum_{ij}\gamma^{\dagger\alpha}_{i\sigma}\gamma_{j\sigma}^{\beta}e^{-i{\bf k(r_i-r_j)}}D_{\alpha\beta}(\tilde{\bf k},\omega),
\end{equation}
where the expansion in powers of the intercluster interaction is implied.
Here $N_c$ is the number of the sites in the cluster, $\gamma_{i\sigma}^{\alpha}=\langle p|d_{i\sigma}|q\rangle$, the wavevector ${\bf k}$ belongs to the original Brillouin zone and ${\bf r_i}$ is the intracluster radius vector.
By construction, the electron Green's function possesses the symmetries of the underlying lattice.

This approach is effectively controlled by the inverse power of the cluster size, $1/N_c$. It is clear that the CPT method is well suited to calculate
properties that are driven by short-range correlations. Long-range correlations are effectively cut off
by the cluster size, and are taken into account only "on average" within a chosen perturbative scheme.
This may result in quantitatively incorrect description of possible phase transitions producing, e.g., an overestimated
value for the corresponding transition temperature. However as soon as the cluster size $N_c$ gets larger the critical temperature goes down, as it should be.

There exists two different approaches to perform the cluster diagonalization within the CPT.
The most frequently used performs calculations with a limited number of low-energy states.
It features the Lanczos algorithm as well as the density matrix renormalization group
method\cite{yang}.
This approach allows one to deal with relatively large clusters ($N_c\geq4\times4$), as long as one discards the high-energy states. Yet this procedure may leads to incorrect results.
The reason for this is that the physics of the lightly doped Mott insulators cannot be separated into low and high energy sectors.
The transitions with relatively high energies may thus have significant spectral weights and if we neglect such contributions it may produce a violation of the total spectral weight. This approach is used in this work to calculate the spectral properties for $4\times 4$ cluster and for $\lambda\sim t$ cases.
Within the second approach employed in this work for $3\times 3$ cluster calculation the whole set of eigenvalues and eigenvectors is calculated entirely.
This kind of diagonalization allows us to control the total spectral weight by taking into explicit account the whole set of possible transitions. However the price we pay for this reflects itself in the fact that the maximal cluster size is significantly reduced in size (the available size of the square clusters are $2\times2$ and $3\times3$).

\begin{figure}[!]
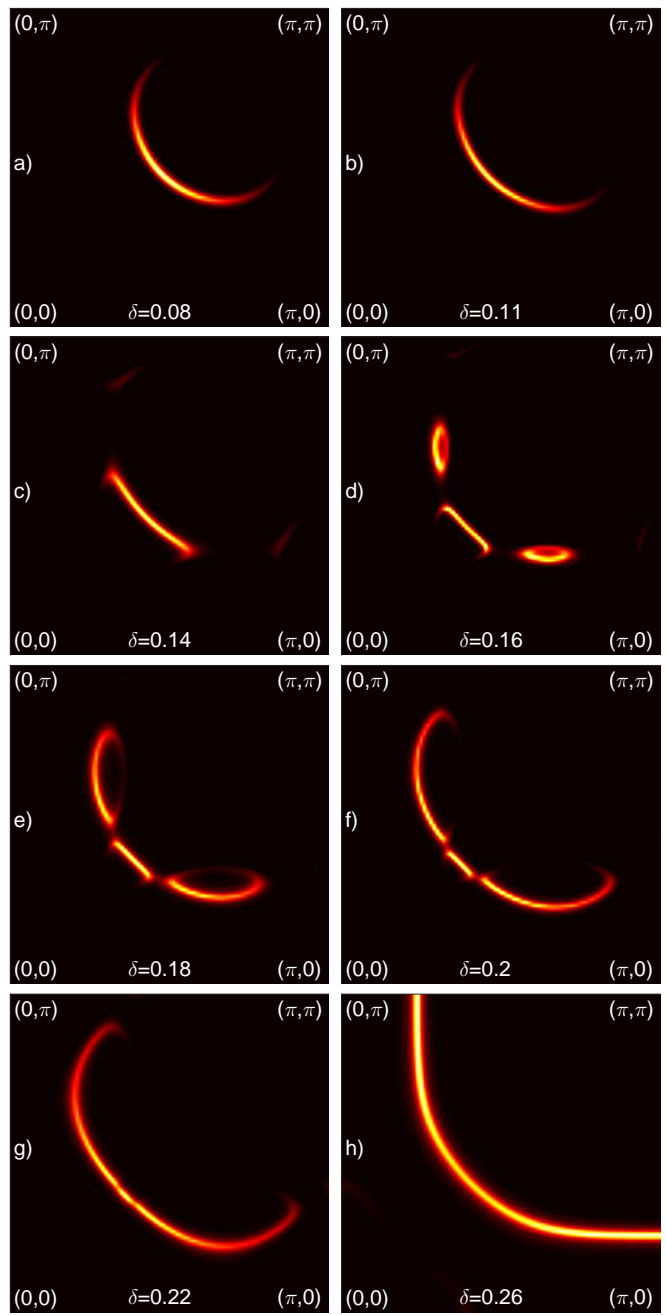

\begin{minipage}[h]{0.49\linewidth}
\center{\includegraphics[width=1\linewidth]{1a.eps}}
\end{minipage}
\hfill
\begin{minipage}[h]{0.49\linewidth}
\center{\includegraphics[width=1\linewidth]{1b.eps}}
\end{minipage}
\vfill
\vspace{0.1cm}
\begin{minipage}[h]{0.49\linewidth}
\center{\includegraphics[width=1\linewidth]{1c.eps}}
\end{minipage}
\hfill
\begin{minipage}[h]{0.49\linewidth}
\center{\includegraphics[width=1\linewidth]{1d.eps}}
\end{minipage}
\vfill
\vspace{0.1cm}
\begin{minipage}[h]{0.49\linewidth}
\center{\includegraphics[width=1\linewidth]{1e.eps}}
\end{minipage}
\hfill
\begin{minipage}[h]{0.49\linewidth}
\center{\includegraphics[width=1\linewidth]{1f.eps}}
\end{minipage}
\vfill
\vspace{0.1cm}
\begin{minipage}[h]{0.49\linewidth}
\center{\includegraphics[width=1\linewidth]{1g.eps}}
\end{minipage}
\hfill
\begin{minipage}[h]{0.49\linewidth}
\center{\includegraphics[width=1\linewidth]{1h.eps}}
\end{minipage}

\caption{The spectral function at the Fermi level in the first quadrant of the BZ is calculated for model (\ref{1.7}) with $J=0.4t$, $\lambda=10^3t$, $T=10^{-3}t$ for different doping levels $\delta$. The second and third nearest neighbor
hopping amplitudes are $t'=-0.27t$ and $t''=0.2t$, respectively. The spectral function is calculated using the CPT method on a $3\times3$ cluster with the Lorentzian broadening $\eta=0.02t$.}
\label{FSd}
\end{figure}

\section{IV. Results}

Our main results are displayed in Fig.\ref{FSd} that reports the actual FS evolution with doping. There are basically four distinct regimes.

i) Small Fermi "arcs" are realized in the nodal region of the BZ for the hole doping level $0.08\le\delta\le0.11$ as depicted in Fig.\ref{FSd}(a-b).
The root cause for these arcs is not entirely clear.
They may presumably occur due to a still unknown translational symmetry breaking order that may set in close to half-filling away from the AF phase.
Such an order is supposed to strongly compete with the superconducting order which results in corresponding short-range correlations.
Short correlation lengths lead to a reduction of the spectral weight\cite{harrison}. Because of this the "arcs" are observed rather than the closed pockets.
In high magnetic fields that annihilates the superconducting state, those "arcs" get closed into small pockets.

ii) An interesting physics is exposed by Figs.\ref{FSd}(c-f). In fact, they display the FS reconstructed by the bidirectional (checkerboard) CDW order.
This is most clearly seen in Fig.\ref{FSd}(d): there is a nodal electron-like Fermi pocket accompanied by two hole-like pockets located closer to the "antinodes".
In this way the reconstructed FS appears as a characteristic feature of the underlying bidirectional CDW order that competes with the superconducting order \cite{sachdev}.
It starts to form at $\delta=0.14$ and it becomes mostly pronounced at $\delta=0.16$. For larger hole concentrations, the hole pockets get enhanced, whereas the electron pocket are reduced in size.
Finally, at $\delta=0.2$, the CDW order terminates. It represents a broken Fermi "arc" as indicated in Fig.\ref{FSd}(f).

Note that one side of the electron pockets depicted in Figs.\ref{FSd}(c-e) is of a rather low intensity. This is due to the fact that in zero magnetic field the CDW order is short-range.
The long-range CDW order is likely to be present only in a strong magnetic field. This is consistent with the observation of well-defined pockets as seen in the quantum oscillation experiments\cite{dlb}. At zero field, the pockets are not well-developed and basically appear as "arcs" as seen in ARPES.

iii) In the doping range $0.2\le \delta\le 0.25$, the CDW order disappears, and the PG phase, with no trace of CDW ordering, is again characterized by the nodal "arc" depicted in Fig.\ref{FSd}(g).

iv)
At hole doping $\delta\ge 0.26$, the large hole-pocket centred at $(\pi,\pi)$ is developed as depicted in Fig.\ref{FSd}(h). Such a FS is typical of conventional FL.
The critical hole concentration that separates large FS from the PG phase is approximately $\delta^{PG}_c=0.25.$

Although the specific values of the hole concentrations that mark different phases are shifted towards relatively larger values compared to those observed in experiments, Fig.\ref{FSd} agrees qualitatively rather well with the experimental data for the underdoped cuprates.
The overestimated values of the critical hole concentrations is presumably an artifact of the finite $3\times 3$ size of the cluster building blocks used in the present CPT theory. Exact diagonalization of larger clusters
poses a well known limitation, however, whereas restricting ourselves to just a set of low-lying excited states results in an uncontrolled error.

Although the conventional $t-J$ model captures the large FS splitting into small pockets/"arcs",
it might seem that the CDW instability represented in Figs.\ref{FSd}(c-f) does not necessarily appear as an intrinsic property of the model.
The point is that, within the CPT approach, we employ the $3\times3$ clusters with the open boundary condition (BC).
This automatically results in an inhomogeneous distribution of the electron density within the isolated cluster.
(Note, however, that the smallest possible $2\times2$ cluster does not result in such an induced CDW order \cite{kuzmin}).

\begin{figure}
\begin{minipage}[h]{0.49\linewidth}
\center{\includegraphics[width=1\linewidth]{2a.eps}}
\end{minipage}
\hfill
\begin{minipage}[h]{0.49\linewidth}
\center{\includegraphics[width=1\linewidth]{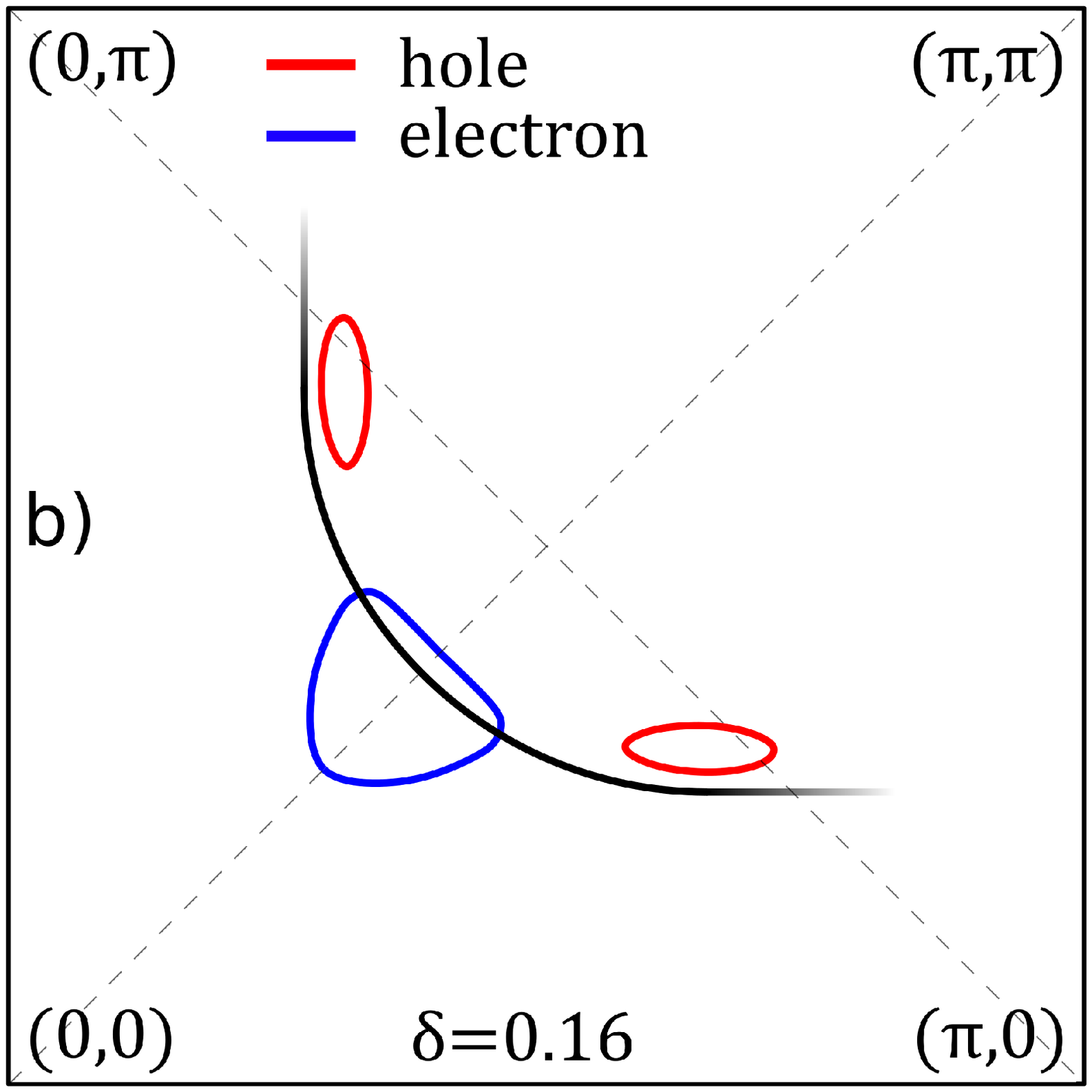}}
\end{minipage}
\caption{Panel (a) shows the Fermi surface for the doping level $\delta=0.16$ (the other parameters are the same as in Fig.\ref{FSd}) with the open BCs being removed by imposing the local chemical potentials (see the main text). Panel (b) schematically shows the Fermi surface at $\delta=0.16$ in the presence of the long-range bidirectional CDW order. The black curve corresponds to the Fermi arc in the absence of the CDW. The red(blue) lines correspond to the poles associated with the hole(electron) pockets which are respectively observed in the presence of the long-range CDW order.}
\label{FS}
\end{figure}

\begin{figure}[!]
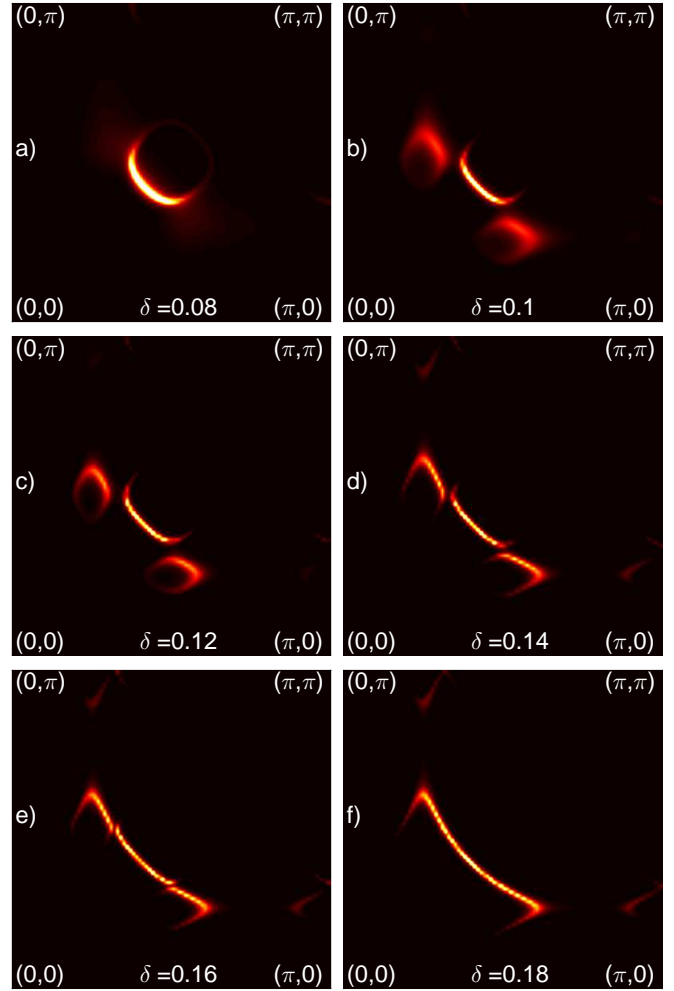

\begin{minipage}[h]{0.49\linewidth}
\center{\includegraphics[width=1\linewidth]{4a.eps}}
\end{minipage}
\hfill
\begin{minipage}[h]{0.49\linewidth}
\center{\includegraphics[width=1\linewidth]{4b.eps}}
\end{minipage}
\vfill
\vspace{0.1cm}
\begin{minipage}[h]{0.49\linewidth}
\center{\includegraphics[width=1\linewidth]{4c.eps}}
\end{minipage}
\hfill
\begin{minipage}[h]{0.49\linewidth}
\center{\includegraphics[width=1\linewidth]{4d.eps}}
\end{minipage}
\vfill
\vspace{0.1cm}
\begin{minipage}[h]{0.49\linewidth}
\center{\includegraphics[width=1\linewidth]{4e.eps}}
\end{minipage}
\hfill
\begin{minipage}[h]{0.49\linewidth}
\center{\includegraphics[width=1\linewidth]{4f.eps}}
\end{minipage}

\caption{The Fermi surfaces calculated by CPT method for the $4\times 4$ cluster. The model parameters are the same as in Fig.\ref{FSd}}
\label{FSd4}
\end{figure}

This is not the case, however.
The CDW order obtained within the $t-J$ model, the main finding of the present paper, is not simply due to the choice of the
$3\times 3$ cluster with the open BCs.
In other words, it is not an artefact of the calculation method but rather the intrinsic property of the $t-J$ model. In fact 
to remove the impact of the open boundary conditions we impose the on-site chemical potentials on each cluster site.
This modification amounts to adding to the Hamiltonian the term:
\begin{equation}
\delta H=\sum_i\mu_in_i,
\end{equation}
where $n_i=\sum_{\sigma}{d}_{i\sigma}^{\dagger} {d}_{i\sigma}$ and the chemical potentials $\mu_i$ are chosen to make the intracluster electron density homogeneous, while the ground state energy is kept intact.
This modification allows us to suppress the intracluster charge density wave generated by the open boundary conditions.
The Fermi surface at the doping level $\delta=0.16$ is calculated using this modification. The result displayed in Fig.\ref{FS}(a) is in qualitative agreement with the FS obtained using the unmodified $3\times 3$ clusters with the open BCs calculation as depicted in Fig.\ref{FSd}(d). This should also be compared to the electron spectral function in the
presence of long-range bidirectional CDW at zero magnetic field as schematically depicted in Fig.\ref{FS}(b).

To explore the effects produced by larger clusters
we derive the FS reconstruction using the $4\times 4$ clusters as building blocks within the employed CPT.
The reconstructed Fermi surfaces are presented in Fig.\ref{FSd4}.
The structure of Fermi surface reconstruction is qualitatively the same as that in the case of $3\times 3$ cluster building blocks. For larger clusters, the values of the critical doping levels of the CDW phase happens to be in a better agreement with the experimental observations.
Our calculation predict the existence of charge density wave in the doping level range $\delta=0.10 - 0.16$.
There are however some distinctions in the location of the hole and the electron pockets.
The reason for this can presumably be traced to the even vs. odd cluster size effects. At the same time, the basic structure of the FS remains qualitatively the same.

\begin{figure}
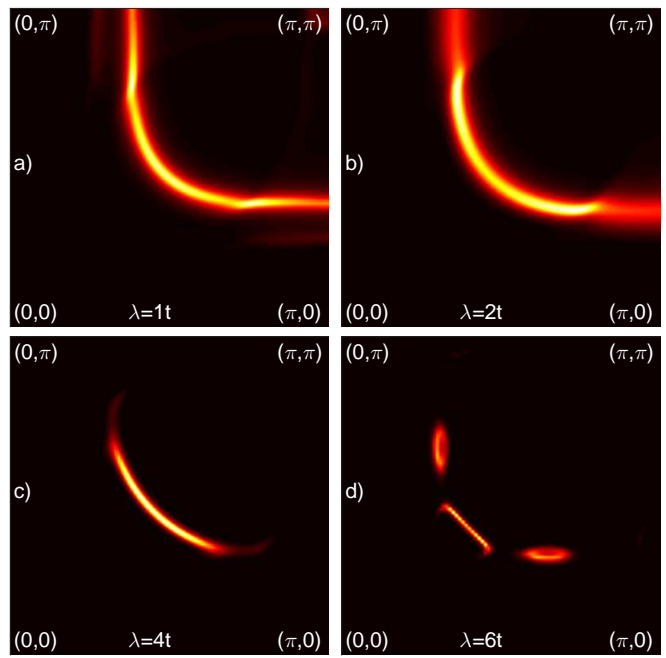

\begin{minipage}[h]{0.49\linewidth}
\center{\includegraphics[width=1\linewidth]{3a.eps}}
\end{minipage}
\hfill
\begin{minipage}[h]{0.49\linewidth}
\center{\includegraphics[width=1\linewidth]{3b.eps}}
\end{minipage}
\vfill
\vspace{0.1cm}
\begin{minipage}[h]{0.49\linewidth}
\center{\includegraphics[width=1\linewidth]{3c.eps}}
\end{minipage}
\hfill
\begin{minipage}[h]{0.49\linewidth}
\center{\includegraphics[width=1\linewidth]{3d.eps}}
\end{minipage}
\caption{The FS with the fixed doping level $\delta=0.16$ and the different $\lambda$ values. Other parameters are the same as in Fig.\ref{FSd}.}
\label{FSl}
\end{figure}

\begin{figure}
\begin{minipage}[h]{0.9\linewidth}
\center{\includegraphics[width=1\linewidth]{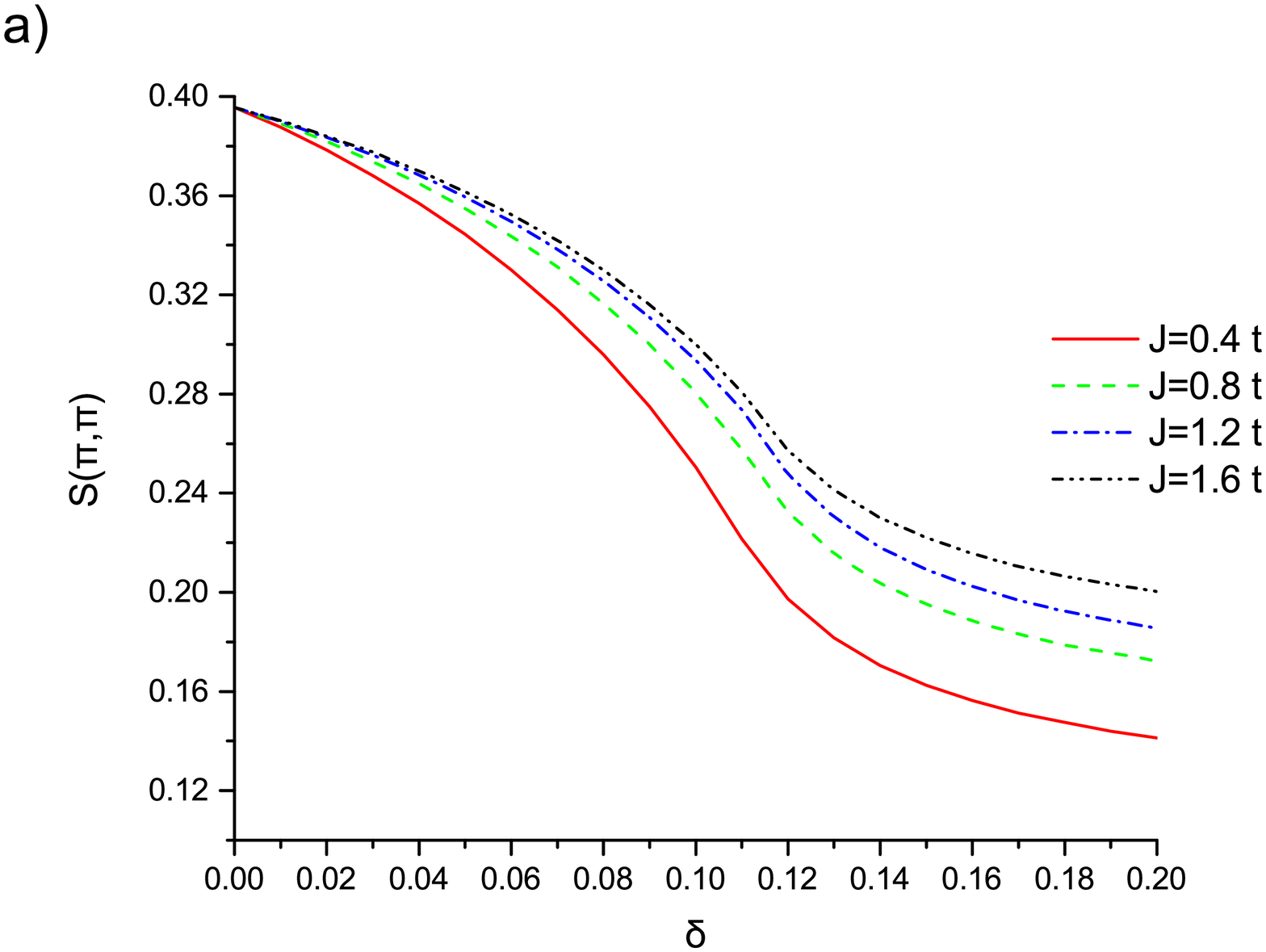}}
\end{minipage}
\vfill
\begin{minipage}[h]{0.9\linewidth}
\center{\includegraphics[width=1\linewidth]{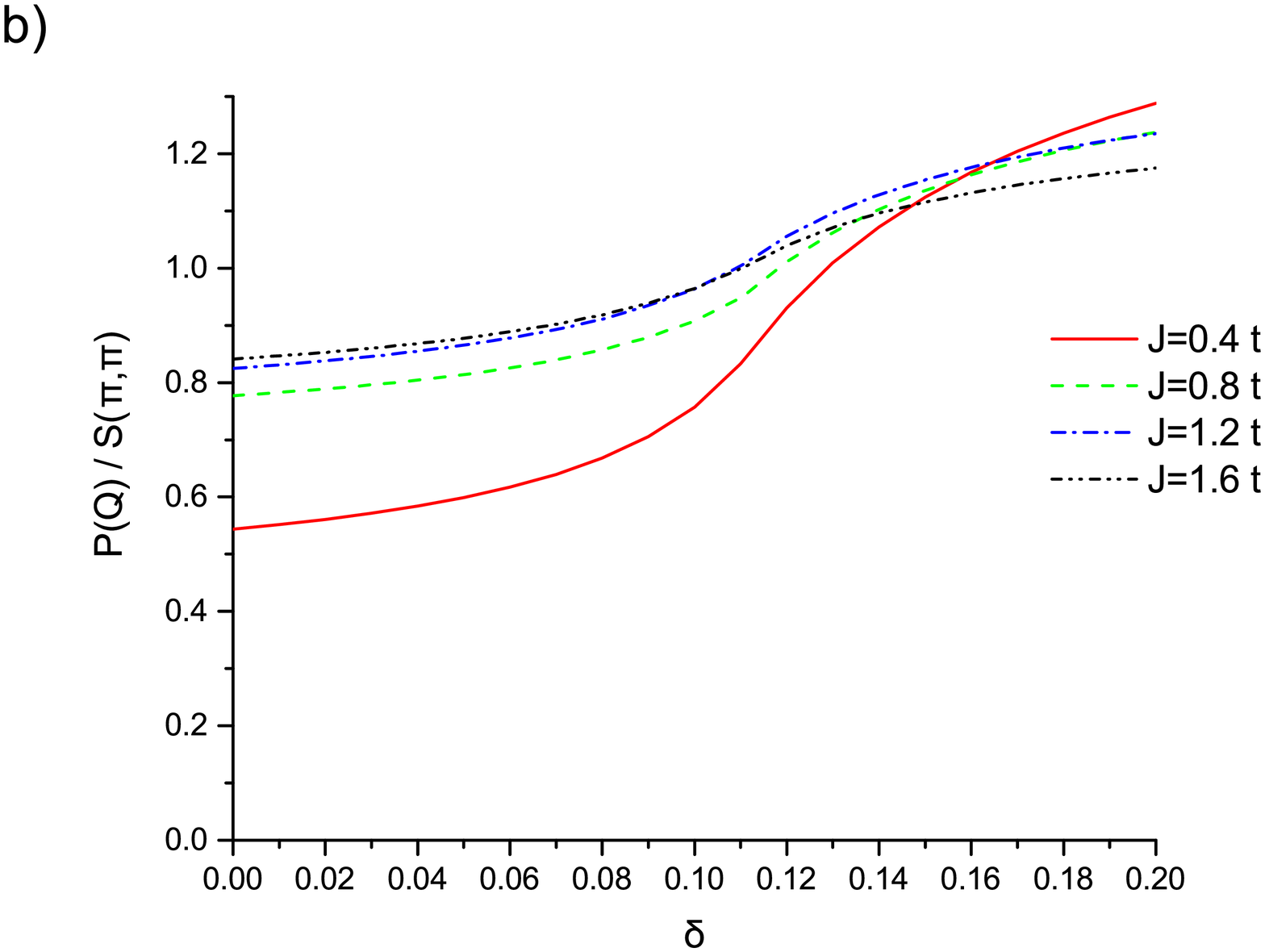}}
\end{minipage}
\caption{Panel (a) shows the dependent of static spin structure factor on doping level $\delta$ for different values of $J$. Panel (b) displays the ratio of the AF structure factor to the CDW structure factor.}
\label{Spp}
\end{figure}

Our model (\ref{1.7}) also allows us to address the issue as to what mechanism actually drives the observed FS reconstruction.
Fig.\ref{FSl} shows that the key point behind the observed FS reconstruction in cuprates is the presence of the strong electron correlations. In the limit of weak correlations ($\lambda < t$), a large Fermi-liquid-like hole FS emerges, whereas for strong coupling ($\lambda=6t$) the reconstructed FS is almost identical in shape to that obtained in the strong coupling limit given by Fig.\ref{FSd}(d). This clearly indicates that the essential physics behind the PG phase is truly driven by strong electron correlations. Note also that although the cluster open BCs are not affected by $\lambda$ the emergent FSs in Fig.\ref{FSl}
clearly display different scenarios, varying between the large hole FS and the induced CDW regime. This observation implicitly indicates that the observed FS reconstruction is a true effect.

It is important to understand why the system develops the CDW correlations instead of the AF correlations
in case the doping level is not too close to the half-filling.
This happens
because of a rapid destruction of the AF long-range order (LRO) in the immediate vicinity of half-filling in the presence of strong electron correlations.
The itinerant-localized duality of the lattice electrons offers the
following explanation of this phenomenon.\cite{ifk2}
The localized individual lattice spins become less correlated
with each other due to the competition between the AF correlations (the characteristic energy scale $\sim J$) and the Kondo screening ($\sim \lambda$) of the local spin moments
by the conduction dopons. The screening breaks the AF bonds.
However if double occupancy is allowed, this breaking is not very efficient, since
it is then induced by a small (in this regime) spin-dopon interaction $\lambda$.
As $\lambda$ increases, the screening becomes more effective.
Since $1/J\gg 1/t$, the hole dynamics is by far the dominant one and it is much faster than the spin dynamics.
The broken AF bonds recover themselves
at a much slower rate than the breaking occurs.
As a result, even a small amount ($\delta \approx 0.03-0.05$) of fast moving dopons (holes) turns out to be, at a large enough $\lambda$, sufficient to completely destroy the AF LRO.

On the other hand, in underdoped cuprates,
the CDW modulations are strongest at $\delta \approx 0.11-0.12$.
At such doping level, the AF LRO is already destroyed by strong correlations so that only
rather slow short-range AF correlations survive.\cite{harrison}
In this parameter range, the AF correlation length turns out to be much smaller than the CDW one.
The latter is noticeably affected only by the superconducting order fluctuations. As soon as those fluctuations are suppressed (e.g., by strong magnetic field) the full CDW long-range ordering sets in as observed in quantum-oscillation experiments.
However, for large enough spin-spin coupling $J$ this picture breaks down as will also be shown in a moment. In this case the AF fluctuations again destroy the CDW order.

To get an insight into in what way the AF and CDW orders compete with each other we first discuss such a competition
within a single 3x3 cluster. In Fig.5(a), we calculate the AF spin density wave (SDW) structure factor
\begin{equation}
S(\pi,\pi)=\frac{4}{N}\sum_{ij}\langle S_i^zS_j^z\rangle e^{i(\vec{r_i}-\vec{r_j})\vec{K}_{\pi\pi}},
\end{equation}
where $\vec{K}_{\pi\pi}=(\pi,\pi)$,
as a function of the hole concentration at various $J$.
The results of the calculation are presented in Fig.\ref{Spp}(a).
Despite the fact that the SDW decreasing rate is underestimated due to the finite cluster size the destruction of the AF order
with doping is clear.

In Fig.\ref{Spp}(b), we compare the AF SDW structure factor with the CDW one.
The charge density wave structure factor reads
\begin{equation}
P(\vec{K})=\frac{1}{N}\sum_{ij\sigma}\langle d_{i\sigma}^{\dagger}d_{j\sigma}\rangle e^{i(\vec{r_i}-\vec{r_j})\vec{K}}.
\label{CDWSF}
\end{equation}
The CDW modulation vectors $Q_x=2\pi(q,0)$ and $Q_y=2\pi(0,q)$, where $q\approx 0.28$ is calculated
within the $3\times 3$ CPT method.
This is in accordance with recent experimental evidence which gives $q\approx 0.3$.\cite{sachdev}
The ratios of $P(\vec{Q})$ to $S(\pi,\pi)$ at the different value of $J$ are presented in Fig.\ref{Spp}(b).
It is clear that at $\delta\approx 0.14$ the CDW order starts to prevail over the SDW one. Right at this doping level
the PG Fermi "arcs" start to reconstruct into the electron/hole pockets induced by the bidirectional CDW.
This observation is in agreement with our $3\times3$ CPT calculations (see Fig.1).

\begin{figure}
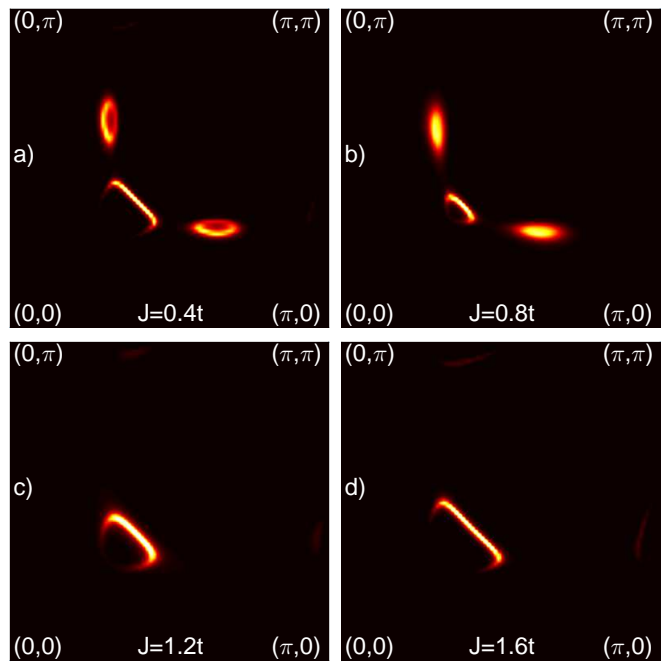

\begin{minipage}[h]{0.49\linewidth}
\center{\includegraphics[width=1\linewidth]{6a.eps}}
\end{minipage}
\hfill
\begin{minipage}[h]{0.49\linewidth}
\center{\includegraphics[width=1\linewidth]{6b.eps}}
\end{minipage}
\vfill
\vspace{0.1cm}
\begin{minipage}[h]{0.49\linewidth}
\center{\includegraphics[width=1\linewidth]{6c.eps}}
\end{minipage}
\hfill
\begin{minipage}[h]{0.49\linewidth}
\center{\includegraphics[width=1\linewidth]{6d.eps}}
\end{minipage}
\caption{The FS with the fixed doping level $\delta=0.16$ and the different $J$ values. Other parameters are the same as in Fig.\ref{FSd}.}
\label{FSJ}
\end{figure}

Finally, within the full lattice CPT, we show that the CDW induced FS reconstruction is destroyed by a large enough spin exchange coupling, $J$, as displayed in Fig.\ref{FSJ}.

\section{V. Conclusion}

In conclusion, the FS reconstruction observed in the underdoped cuprates can be accounted for within a conventional microscopic $t-J$ model. This reconstruction is clearly induced by strong electron correlations. The bidirectional CDW instability observed in several recent experiments is also present in this model.
Our work provides strong arguments to believe that these results are not artefacts of small lattices, but rather represent true physical effects due to strong electron-electron correlations.

\begin{acknowledgments}

The authors thank Marcin Mierzejewski, Dionys Baeriswyl and Catherine Pepin for stimulating discussions.
We would like to thank as well an anonymous referee for stimulating critical remarks.
\end{acknowledgments}

\end{document}